\documentclass[pdftex,twocolumn,epjc3]{svjour3}          

\RequirePackage[T1]{fontenc}

\smartqed  

\usepackage{tikz-feynman}
\RequirePackage{graphicx}
\RequirePackage{mathptmx}      
\RequirePackage{flushend}
\RequirePackage[numbers,sort&compress]{natbib}
\RequirePackage[colorlinks,citecolor=blue,urlcolor=blue,linkcolor=blue]{hyperref}

\journalname{Eur. Phys. J. C}

\begin{document}

\title{Searching for Dark Matter with $t \bar{t}$ Resonance}

\author{Yoav Afik\thanksref{e1,addr1}
        \and
        Eitan Gozani\thanksref{e2,addr1} 
        \and
        Yoram Rozen\thanksref{e3,addr1} 
}

\thankstext{e1}{yoavafik@campus.technion.ac.il}
\thankstext{e2}{gozani@campus.technion.ac.il}
\thankstext{e3}{rozen@physics.technion.ac.il}

\institute{Department of Physics, Technion: Israel Institute of Technology\\ Haifa, Israel\label{addr1}
}

\date{Received: date / Accepted: date}

\maketitle

\begin{abstract}
Many models containing particles which are candidates for dark matter, assume the standard model particles and the dark matter candidates are mediated by a spin-0 particle.
At the LHC, one can use these models for dark matter searches.
One of the possible approaches for the search of these models is by considering the decay of the spin-0 particle to a pair of $t\bar{t}$, thus modifying the pattern of the top quark pair invariant mass spectrum.
This search suggests a good sensitivity in a parameter space different than the more traditional searches.
We examine this  sensitivity and put limits on two benchmark models containing  candidates for dark matter, using previous ATLAS results. It was found that when the mediator mass ($m_{Y_0}$) and the dark matter candidate mass ($m_{\chi}$) have values of $m_{Y_0} \sim 2 \cdot m_{\chi}$, mediator masses in the range of $[400,600]$ GeV are excluded.
We compare our results to direct detection experiments and show that we gain sensitivity for new regions which are not covered by other searches.
\end{abstract}

\section{Introduction}
\label{sec:intro}
Astrophysical observations support the existence of nonbaryonic component of the universe:
Dark Matter (DM) \cite{Zwicky:1933gu, DMreview, Jungman:1995df, Binney:1993ce}. 
DM particles have to be stable, massive, and do not participate in the strong and electro-magnetic interactions.
There are many searches for DM candidates at the Large Hadron Collider (LHC) experiment
that use different approaches to model the signal for DM.
One of the most popular candidates is a Weakly Interacting Massive Particle (WIMP) \cite{Steigman:1984ac}.
At the LHC, one can search for WIMP type DM particles ($\chi$) produced in $pp$ collisions.

Searches for DM using models with DM candidates and spin-0 mediators as a signal were already presented by ATLAS \cite{ATLAS-CONF-2016-077, ATLAS-CONF-2016-050, ATLAS-CONF-2016-076, ATLAS-CONF-2016-086, Aaboud:2017rzf, ATLAS-CONF-2018-051} and CMS \cite{Sirunyan:2017xgm, CMS-PAS-EXO-16-005, CMS-PAS-EXO-16-028, CMS-PAS-B2G-15-007, Sirunyan:2018dub} collaborations, with up to $36 fb^{-1}$ of integrated luminosity, with centre of mass energy $\sqrt{s}=13$ TeV. These searches focus on production of DM in association with a pair of top or bottoms quarks. 
In all of these searches, the mediator decays to a pair of DM candidates, leaving a signature of high missing transverse momentum in the detector.

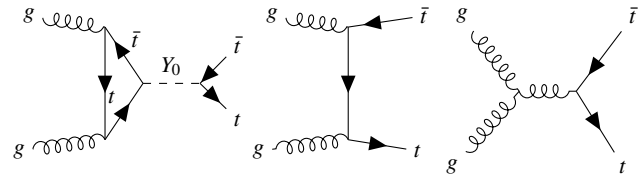
\begin{figure}
  \centering
  
\begin{tikzpicture}
  \begin{feynman}
    \vertex (a1) {$g$};
    \vertex[right=1cm of a1] (a2);
    \vertex[right=1cm of a2] (a3);
    \vertex[below=6em of a3] (b1);
    \vertex[left=1cm of b1] (b2);
    \vertex[left=0.95cm of b2] (b3) {$g$};
    \vertex[below=3.0em of a2] (c1);
    \vertex[right=0.5cm of c1] (c2);
    \vertex[below=0.5em of a2] (d1);
    \vertex[above=0.5em of b2] (d2);
    \vertex[right=0.75cm of c2] (e1);
    \vertex[right=0.5cm of e1] (e2);
    \vertex[above=1.0em of e2] (e3) {$\bar{t}$};
    \vertex[below=1.0em of e2] (e4) {$t$};

    \vertex[right=3.2cm of a1] (f1) {$g$};
    \vertex[right=1cm of f1] (f2);
    \vertex[right=1cm of f2] (f3);
    \vertex[below=6em of f3] (g1);
    \vertex[left=1cm of g1] (g2);
    \vertex[left=1.0cm of g2] (g3) {$g$};
    \vertex[below=3.0em of f2] (h1);
    \vertex[right=0.5cm of h1] (h2);
    \vertex[below=0.5em of f2] (i1);
    \vertex[above=0.5em of g2] (i2);
    \vertex[right=2.0cm of f1] (j1) {$\bar{t}$};
    \vertex[right=2.1cm of g3] (j2) {$t$};

    \vertex[right=5.7cm of a1] (k1) {$g$};
    \vertex[right=0.75cm of k1] (k2);
    \vertex[below=1.0cm of k2] (l1);
    \vertex[below=1.0cm of l1] (l2);
    \vertex[left=0.7cm of l2] (k3) {$g$};
    \vertex[right=0.75cm of l1] (m1);
    \vertex[right=2.25cm of k1] (n1){$\bar{t}$};
    \vertex[right=2.25cm of k3] (n2){$t$};

    \diagram* {
      {[edges=fermion]
      },
      (a1) -- [gluon] (d1),
      (d1) -- [fermion] (d2),
      (d2) -- [gluon] (b3),
      (d1) -- [anti fermion, edge label=\(\bar t \)] (c2),
      (d2) -- [fermion, edge label=\(t \)] (c2),
      (c2) -- [scalar, edge label=\(Y_0\)] (e1),
      (e1) -- [anti fermion] (e3),
      (e1) -- [fermion] (e4),
      
      (f1) -- [gluon] (i1),
      (i1) -- [fermion] (i2),
      (i2) -- [gluon] (g3),
      (i1) -- [anti fermion] (j1),
      (i2) -- [fermion] (j2),

      (k1) -- [gluon] (l1),
      (k3) -- [gluon] (l1),
      (l1) -- [gluon] (m1),
      (m1) -- [anti fermion] (n1),
      (m1) -- [fermion] (n2),

    };

  \end{feynman}
\end{tikzpicture}

	    \caption{Representative Feynman diagrams for a $t \overline{t}$ production via spin-0 mediator ($Y_0$) and via the SM.}
	    \label{fig:Feynman1}
\end{figure}

When searching for dark matter, most of the analyses use some model as a benchmark.
Although the conclusions of a given analysis are broader than the benchmark model used, in most of the cases, the analysis aims to maximise the sensitivity for this specific model.
As general as the model is, parameter spaces of other models containing DM candidates are not necessarily covered. 
A complementary search for these models can be achieved if the mediator decays to a pair of top quarks, leaving a more complex signature in the detector.
These searches are challenging as strong interference with the Standard Model (SM) $t \bar{t}$ production is expected \cite{Hespel:2016qaf}, leading to a pick-dip shape in the spectrum of the $t \bar{t}$ invariant mass \cite{Dicus:1994bm, Frederix:2007gi}.
Good sensitivity is expected  if the mediator is heavier than twice the mass of the top quark.
Some representative Feynman diagrams for leading-order production of a $t \bar{t}$ pair by a spin-0 mediator ($Y_0$) and by the SM are presented in Figure \ref{fig:Feynman1}.

There have been a few analyses targeting searches for heavy particles decaying to a pair of top quarks \cite{Chatrchyan:2012ku, Aad:2015fna, Aaboud:2018mjh}, creating a Breit-Wigner resonance in the $m_{t \bar{t}}$ spectrum.
However, for most of those analyses, spin-0 particles were not taken into account. 
The recent search published in a similar context, considering interference with the SM for $t \bar{t}$ production, was done by the ATLAS collaboration \cite{Aaboud:2017hnm} using data of $pp$ collisions at a centre-of-mass energy of $\sqrt{s}=8$ TeV and integrated luminosity of $20.3 fb^{-1}$.
This analysis used a Two-Higgs-Doublet model for the interpretation of the results, with a spin-0 particle mass of $500-750$ GeV.
Here, we show how the search for $t \bar{t}$ resonance originating from spin-0 particles is important for models containing DM. We cover parameter space that is not covered by the more traditional searches. Especially, searches for signature with a mediator that decays to a pair of DM candidates.


\section{Theoretical Framework}
\label{sec:framework}

In many models containing  new spin-0 particles, the couplings with the SM fermions are being set proportional to the SM Yukawa terms, by using the Minimal Flavour Violation (MFV) assumption \cite{DAmbrosio:2002ex}.
This  motivates  a search in association with heavy flavour quarks.
There are many models that assume interactions between DM candidates and spin-0 CP-odd or CP-even mediators, see for example \cite{Buckley:2014fba, Berlin:2014cfa, Bauer:2017ota, Bell:2017rgi, Arcadi:2018pfo, Han:2018bni, Bandyopadhyay:2017tlq, Dutta:2017jfj}.

The width, at tree level, 
for a spin-0 particle decaying to a pair of Dirac fermions, which can be either DM candidates ($\chi \bar{\chi}$) or SM fermions ($f \bar{f}$), for models assuming MFV,
is calculated  as follows:
\begin{equation}
\Gamma(\phi / a \rightarrow \chi \bar{\chi}) = (g_{med-\chi \bar{\chi}})^2 \cdot \frac{m_{\phi / a}}{8\pi}\Big(1-\frac{4 \cdot m_{\chi}^2}{m_{\phi / a}^2}\Big)^{n/2}
\label{eq:Width_chichi},
\end{equation}
\begin{equation}
\Gamma(\phi / a \rightarrow f \bar{f}) = (g_{med-f \bar{f}})^2 \cdot \frac{y_f^2 \cdot m_{\phi / a}}{16\pi} \Big(1-\frac{4 \cdot m_{f}^2}{m_{\phi / a}^2}\Big)^{n/2}
\label{eq:Width_ff},
\end{equation}
where $n=3$ for a scalar ($\phi$) and $n=1$ for a pseudo-scalar ($a$). 
Here, $\chi$ is the DM candidate, $f$ is the SM fermion, $m_{\phi / a}$ is the mass of the scalar / pseudo-scalar, $m_\chi$ is the DM mass, $m_f$ and $y_f$ are the corresponding mass and Yukawa term for the SM fermion, respectively. The parameters $g_{med-\chi \bar{\chi}}$ and $g_{med-f \bar{f}}$ are model dependent couplings.
In general, equation \ref{eq:Width_chichi} can be applied to other types of DM candidates, which are not Dirac fermions, but we keep it as a benchmark assumption.
Interactions between the dark sector and the SM gauge bosons exist in part of those models, and are taken into account when analysing the results. 

\begin{figure}
\centering
\includegraphics[width=.50\textwidth]{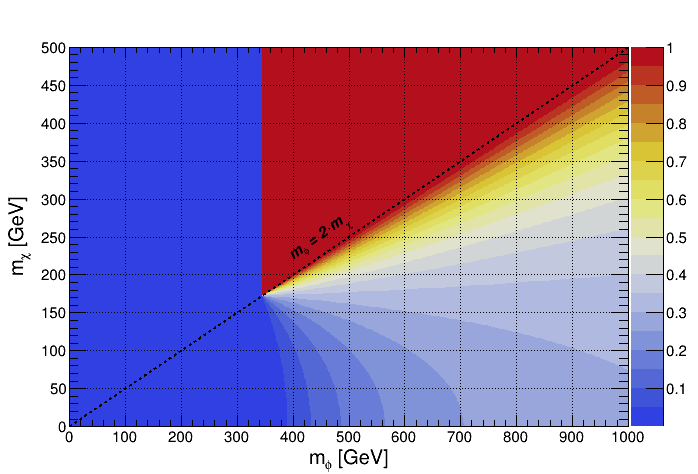}
\includegraphics[width=.50\textwidth]{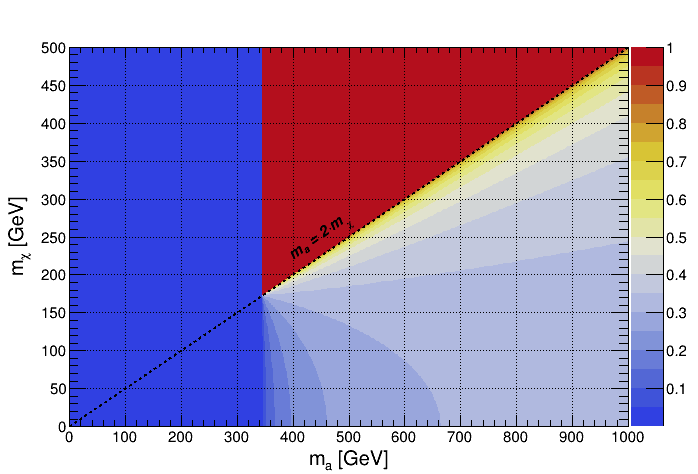}

\caption{Branching Ratio for the decay of a spin-0 mediator to a pair of $t \bar{t}$, assuming interactions with dark matter and tops only. 
The Branching Ratio is presented for a scalar (up) and pseudo-scalar (bottom) scenarios.
The diagonal black dashed lines present the limit in the parameter space where the mediator is heavy enough to decay to a pair of DM candidates. 
}
\centering
\label{fig:widths}
\end{figure}

The calculation of the mediator width from equations \ref{eq:Width_chichi} and \ref{eq:Width_ff} presents an interesting behavior:
If the mediator is heavy enough to decay to a pair of top quarks (i.e. $f \bar{f} = t \bar{t}$), the partial decay width of $\phi / a \rightarrow \chi \bar{\chi}$ becomes significantly smaller. This is especially true for high $m_\chi$, where the partial decay width of $\phi \rightarrow \chi \bar{\chi}$ is highly suppressed.
Therefore, the $t \bar{t}$ resonance search along the search for $t \bar{t} + \chi \bar{\chi}$ are complementary: the former gains better sensitivity for low DM masses, while the latter has a better sensitivity for high DM masses.
The Branching Ratio of this kind of spin-0 decay to a pair of top quarks is presented in Figure \ref{fig:widths}, assuming couplings only for DM and top quarks, and setting $g_{med-\chi \bar{\chi}} = g_{med-f \bar{f}} = 1$.
The behaviour discussed above is well observed in those figures.

\section{Benchmark Models}
\label{sec:benchmarks}

In order to emphasize how the behavior described in section \ref{sec:framework} affects the sensitivity for models containing a spin-0 mediator and DM candidates, we select two suitable benchmark models.
The first model we consider is a simplified model with a spin-0 mediator that couples to both the SM fermions and to a new dark sector \cite{Abercrombie:2015wmb,Buckley:2014fba,Haisch:2015ioa}, where the couplings of the spin-0 mediator and the SM fermions are universal.
The second model is a type-II two-Higgs-doublet model with an additional U(1) gauge symmetry, introduced in \cite{Berlin:2014cfa}.

The first model describes a general set of couplings, and assumes that only fermions interact with the spin-0 mediators.  
The second model however, also describes interactions between the SM bosons and the spin-0 mediator, affecting the width of the spin-0 particle, and therefore modifying the invariant mass spectrum of the top quark pair.
In addition, in the second model, unlike the first one, the couplings of the spin-0 mediator to the SM fermions are not universal, and therefore are different between down-type and up-type quarks (they are all fixed by one parameter, but they have different values).

\subsection{Benchmark 1: Simplified Models}

We consider two choices: in the first one the interaction with the SM is mediated by a scalar,
while in the second scenario we consider only a new pseudo-scalar (assuming that the associated scalar is decoupled from the low-energy spectrum). 
The dark sector, in general, can contain more than a single particle. We assume the dark sector contain only one DM candidate which is a Dirac fermion. 
We keep in mind this assumption affects mostly the width of the mediator, so the results can be converted easily to more complicated cases.
This model assumes Yukawa-like couplings between the dark sector
mediator and the SM fermions.
The interaction terms of the Lagrangians for the scalar $({\mathcal L}_{\phi})$ and pseudo-scalar $({\mathcal L}_{a})$ are \cite{Buckley:2014fba}:
\begin{equation}
\mathcal{L}_{\phi}^{int} = - g_{\chi} \phi \overline{\chi} \chi - \sum_{fermions} g_{v} \frac{y_{f}}{\sqrt{2}} \phi \overline{f} f 
\label{eq:scalar_lagrangian},
\end{equation} 
\begin{equation}
\mathcal{L}_{a}^{int} = - ig_{\chi} a\overline{\chi} \gamma^{5} \chi - \sum_{fermions} ig_{v} \frac{y_{f}}{\sqrt{2}} a \overline{f} \gamma^{5} f 
\label{eq:pseudo-scalar_lagrangian}.
\end{equation} 
Here, $\phi$ and $a$ are scalar and pseudo-scalar fields which connect the SM with the dark sector,
$\chi$ is the DM field, $g_{\chi}$ is the DM-mediator coupling, $g_{v}$ is the flavour-universal SM-mediator coupling and $y_{f}$ are the SM Yukawa couplings for fermions.

If $g_{v} \sim g_{\chi}$ and $m_{\phi/a} > 2 m_{\chi}$, the decay of the mediator to DM is expected to dominate, unless the mediator is heavy enough for the top channel to open.
This holds true because the Yukawa couplings to light fermions are significantly smaller compared to the Yukawa term of the top.
The minimal viable value of $\Gamma_{\phi/a}$ can be calculated from the
other parameters. The mediator width can 
be larger  than the minimal one if additional dark sector
particles are present. In our interpretation, however, we assume only one type of DM candidate. 
For simplicity, 
we use $g = g_\chi = g_v$,
reducing  the  free parameters to three: $m_{\chi}$, $m_{\phi/a}$, $g$.
To simulate the signal, the
 {\sc DMSimp} \cite{Mattelaer:2015haa,Backovic:2015soa,Afik:2018rxl} models have been used with MadGraph \cite{Alwall:2011mad}.

\subsection{Benchmark 2: 2HDM+$Z'$}

This model is an extension to the familiar type-II 2HDM model \cite{Branco:2011iw}, and introduces an extra spin-1 mass eigenstate which is denoted as $Z'$.
The pseudo-scalar ($A_0$) is the only one that couples to a pair of DM candidates, therefore this particle is identified as the mediator between the SM particles and the dark sector. The decoupling limit \cite{Gunion:2002zf} is assumed.
Limits were set recently on this model by ATLAS  \cite{ATLAS-CONF-2018-051}, using final state with the SM Higgs boson (decays to a pair of bottom quarks or photons) and $A_0$ (decays to a pair of DM candidates). However, the parameter choice was set in order to emphasize the sensitivity of the signatures above. Therefore, we choose different parameters, in order to emphasize the final state discussed in this paper, and to add new constraints on the model.

This model has six parameters, which are set as follows:
The mass of the light scalar, recognised as the SM Higgs boson in the decoupling limit, is set to $m_h$ = 125 GeV;
In order to to avoid interference effects between two bosons, we choose the masses of the spin-0 bosons to allow only one of them, $A_0$, to decay to a $t \bar{t}$ pair. Therefore, the mass of the heavier scalar and charged scalar are set equal to each other, $m_H = m_{H^\pm}$ = 300 GeV, and the mass of the pseudo-scalar is set to $m_{A_0}$ = 400 GeV; 
The parameters of the new $Z'$ boson, it's mass ($m_{Z'}$) and coupling  ($g_{Z'}$), has a negligible effect on the $t \bar{t}$ invariant mass spectrum, and are chosen randomly to be
$m_{Z'}$ = 3 TeV and $g_{Z'}$ = 1. The results are valid for other values of $m_{Z'}$, as long as it is heavy enough to avoid decays of $A_0$ to $Z'$ with another Higgs boson; 
The ratio between vacuum expectation values of the two Higgs doublets, $tan(\beta)$, and the mass of the DM candidate, $m_\chi$, are free parameters.
Those two parameters were selected to be scanned, since they have a direct impact on the strength of $A_0$ interactions and decay width, and therefore on the top pair invariant mass spectrum.


\section{Results}
\label{sec:results}

Limits on spin-0 mediator models were already set at 95\% Confidence Level (CL) \cite{Hespel:2016qaf}, using the latest ATLAS $t \bar{t}$ resonance search with available data \cite{Aad:2015fna}. 
ATLAS  also set limits on spin-0 mediators \cite{Aaboud:2017hnm}.
In both cases interference effects with the SM were considered, and the signal was modeled at Next to Leading Order (NLO) and Next to Next to Leading Order (NNLO) in QCD corrections, respectively.
For those limits, however, no interaction with DM candidates was taken into account. This type of interaction is being considered in this section.
The results of the former were found to be more efficient for mediator masses which are lower than 500 GeV, while the latter is more efficient for mediator masses which are higher than 500 GeV.
This is expected since the corresponding ATLAS analysis targeted spin-0 particle masses which are higher than 500 GeV.

The experimental resolution on the $t \bar{t}$ invariant mass, $m_{t \bar{t}}$, was calculated to be 8\% for both analyses.
Since the width of the mediator has a strong effect on the shape of the pure signal and interference distributions, an upper limit $\frac{\Gamma_{total}}{m_{\phi / a}} < 8\%$ was set, where $\Gamma_{total}$ is the total decay width of the mediator.
In the case that the mediator decays to a pair of DM candidates, $\frac{\Gamma_{total}}{m_{\phi / a}} < 40\%$ was used to keep the narrow width approximation valid.
Results with higher total widths were discarded.

\subsection{Benchmark 1: Simplified Models}
\label{sec:results_simplified}

\begin{figure}
\centering
\includegraphics[width=.50\textwidth]{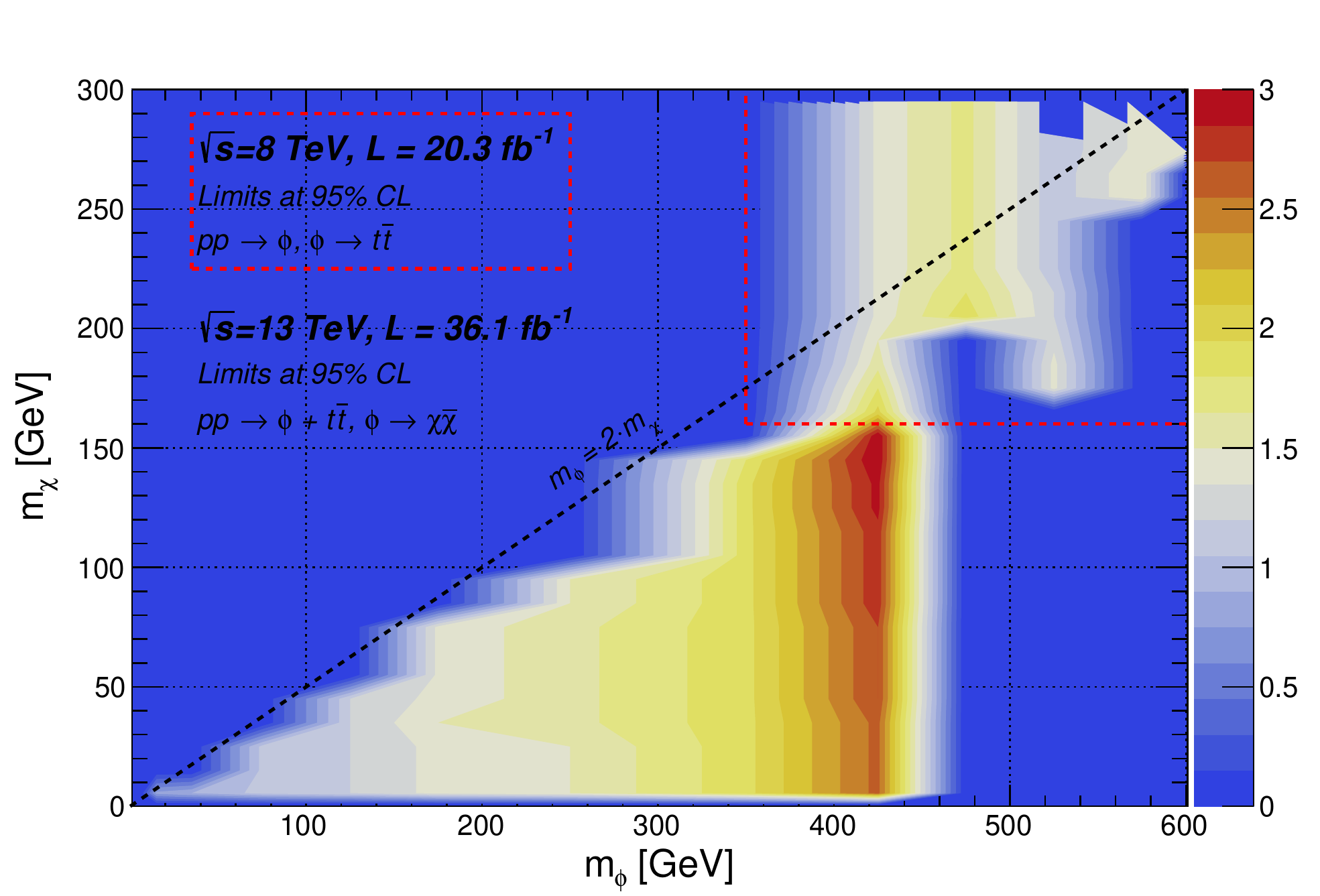}
\includegraphics[width=.50\textwidth]{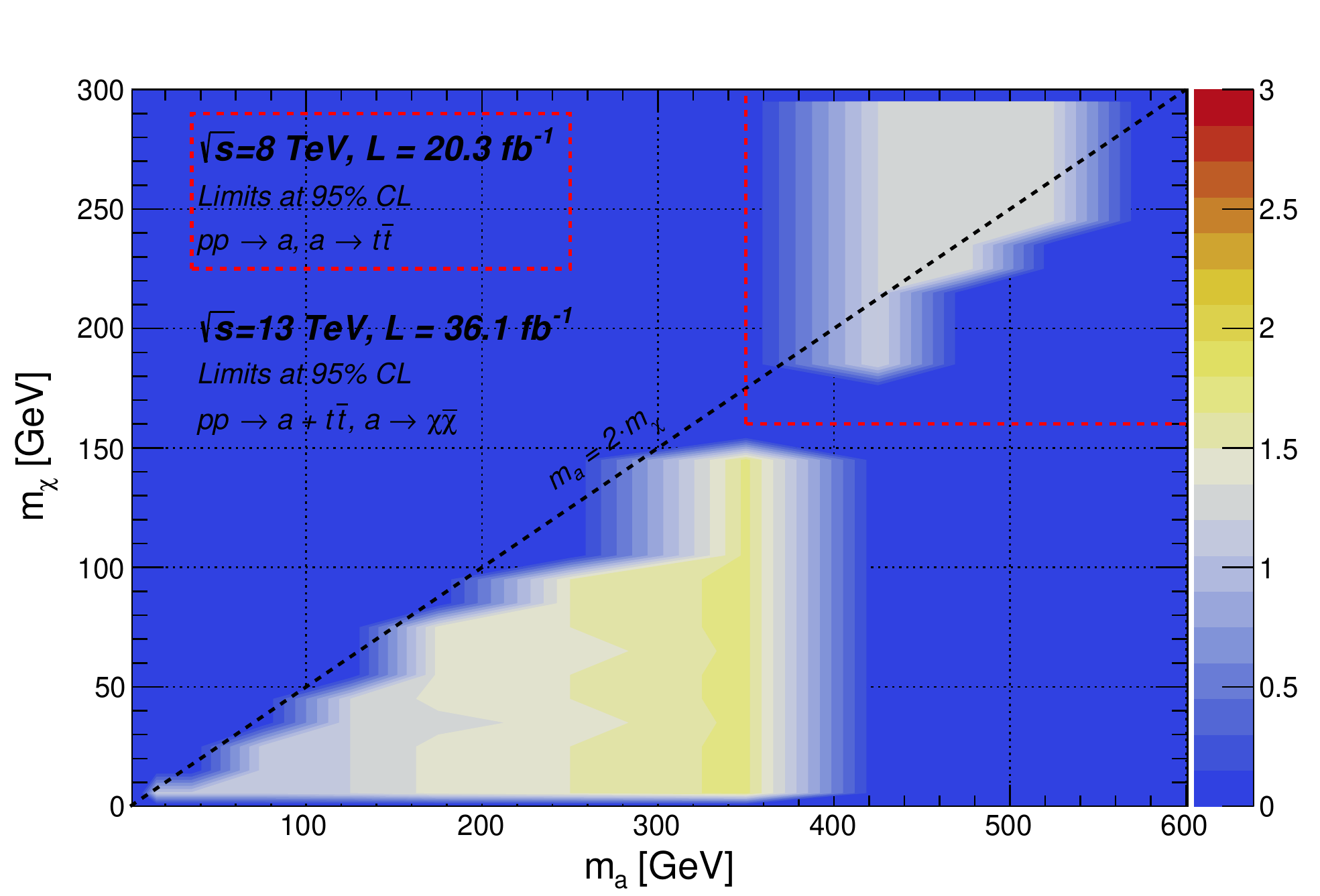}

\caption{Exclusion contour for DM simplified models in the $(m_{\phi / a}, m_{\chi})$ plane for the scalar (upper) and pseudo-scalar (bottom) scenarios.
The contour corresponds to the lowest value of the coupling $g = g_\chi = g_v$ allowed. Results from both $t \bar {t}$ resonance and $t \bar{t} + \chi \bar{\chi}$ signatures are presented for comparison, while the red dashed lines delimitate the limits set by the $t \bar {t}$ resonance signature.
The diagonal black dashed lines present the limit in the parameter space where the mediator is heavy enough to decay to a pair of DM candidates. 
}
\centering
\label{fig:upper_limit_simp}
\end{figure}

Figure \ref{fig:upper_limit_simp} presents upper limits at 95\% CL on the coupling $g$.
The figure presents the lowest coupling excluded for the model. 
Both scalar and pseudo-scalar mediator cases are considered. 
The best limits obtained from \cite{Aaboud:2017rzf} searching for $t \bar{t}+\chi \bar{\chi}$ processes are presented for comparison.
The exclusion contour is more stringent for the $t \bar{t}$ resonance searches when $m_{\phi / a} \geq 400$ GeV, especially when the DM mass is high.
The limits obtained from the $t \bar{t}$ resonance are stronger for the scalar case since the width calculation (see equations \ref{eq:Width_chichi}, \ref{eq:Width_ff}) allows higher values for the pseudo-scalar case with similar parameters, leading to higher total widths which we discard.

The areas above the diagonal black dashed lines at Figure \ref{fig:upper_limit_simp} present the parameter space where the mediator is virtual for the production of $t \bar{t}+\chi \bar{\chi}$ processes ($m_{\phi / a} < 2 \cdot m_{\chi}$).
In this part of the parameter space, comparing to the on-shell case (the area below the line with $m_{\phi / a} > 2 \cdot m_{\chi}$), the sensitivity for $t \bar{t}+\chi \bar{\chi}$ processes decreases significantly.
This happens mainly because the cross section is lower (see for example Figure 6 at \cite{Aaboud:2017rzf}).

The regions of the parameter space where the mediator mass is higher than twice the mass of the top ($m_{\phi / a} > 2 \cdot m_{t}$) are covered by the signature discussed in this paper for two cases: 
The first case is above the diagonal line, where the mass of the DM does not play a significant role, since in this case the mediator width is completely dominated by the top pair decay;
the second case is below the diagonal line, allowing on-shell decay of the mediator both to a pair of top quarks and a pair of DM candidates.
The limits in the second case are set close to the diagonal line, since the partial decay width of the mediator to $\chi \bar{\chi}$ decreases when $2 \cdot m_{\chi}$ increases up to $m_{\phi / a}$, as concluded from Equation \ref{eq:Width_chichi}.

For both of the regions in the parameter space discussed in the previous paragraph, the $t \bar{t}+\chi \bar{\chi}$ search is not sensitive. However, the signature discussed in this paper is sensitive.
On the contrary, in the case that the mediator is not heavy enough to decay to a pair of tops ($m_{\phi / a} < 2 \cdot m_{t}$) but decays to a pair of DM candidates ($m_{\phi / a} > 2 \cdot m_{\chi}$), the $t \bar{t}+\chi \bar{\chi}$ search is sensitive while the final state discussed in this paper is not.
This emphasize the necessity of those signatures as being complementary to each other for this model.


\subsection{Benchmark 2: 2HDM+$Z'$}
\begin{figure}
\centering
\includegraphics[width=.50\textwidth]{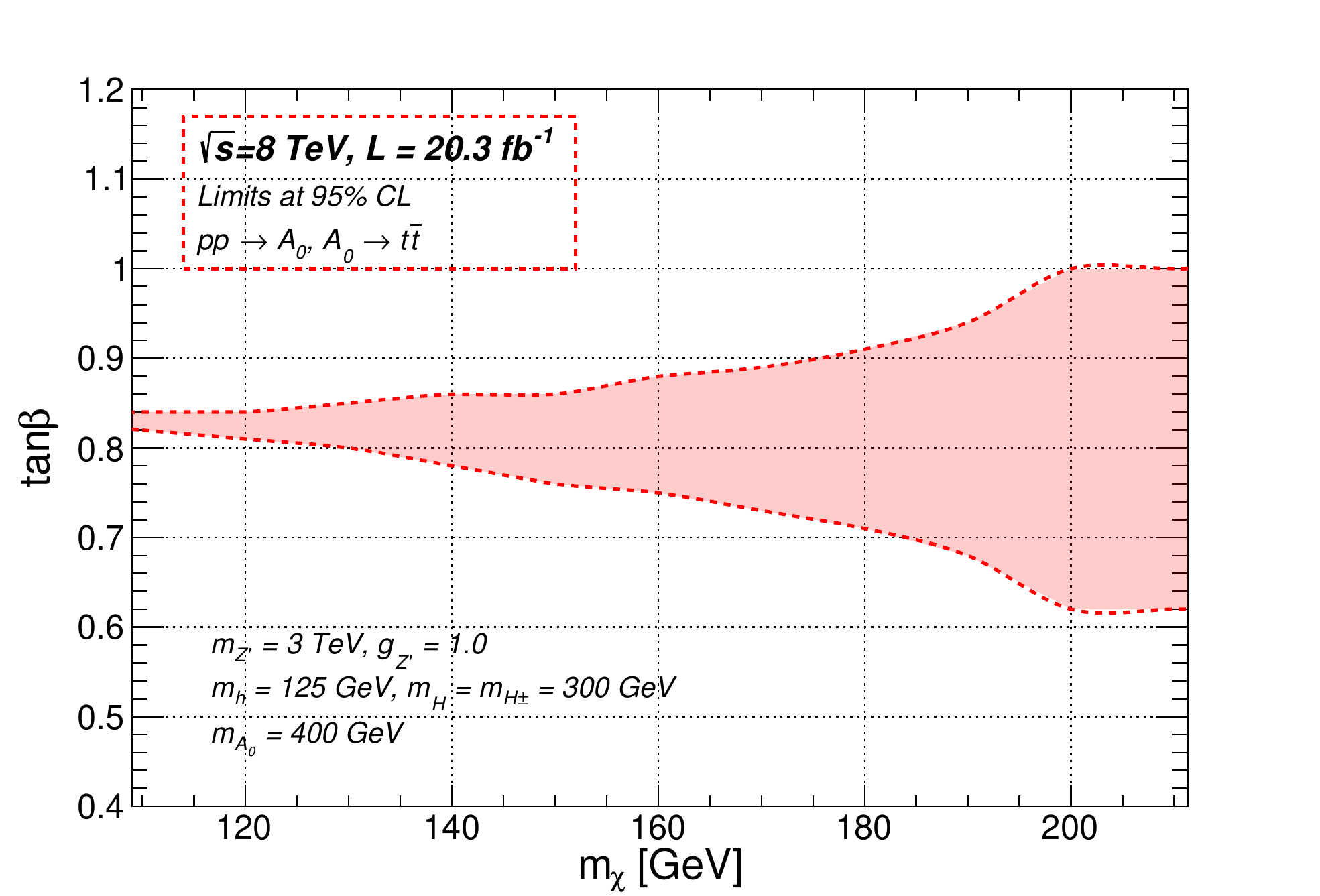}

\caption{Exclusion contour for 2HDM+$Z'$ model in the $(m_{\chi},tan \beta)$ plane. The contour corresponds to the excluded values of the model. All of the other parameters of the model are fixed and mentioned in the figure. Only results using the $t \bar {t}$ resonance signature are presented, since other signatures do not exclude any part of the presented parameter space.
}
\centering
\label{fig:upper_limit_2HDM_Z}
\end{figure}

Figure \ref{fig:upper_limit_2HDM_Z} presents the exclusion contour at 95\% CL in the plane of $(m_\chi,tan \beta)$.
The remaining  parameters of the model are fixed as discussed in section \ref{sec:benchmarks}.
The best limits obtained from other searches are not presented since they do not provide any constraint in the considered parameter space.
In this benchmark model the coupling between the top quarks and $A_0$ is proportional to $(tan \beta)^{-1}$. Therefore, the lower part of the exclusion contour is due to mediator width values which are higher than 8\% (see equation \ref{eq:Width_ff} with $g_{med-f \bar{f}} = (tan \beta)^{-1}$), while the upper part is due to lower cross sections not excluded by the analysis.

\subsection{Comparison to Direct Detection Experiments}
Results from LHC analyses with DM simplified models can be compared to direct detection experiments, for both spin-independent and spin-dependent cases.
For this purpose, we use the prescription described at \cite{Boveia:2016mrp}, setting limits on DM-nucleon interaction.

\subsubsection{Spin-Independent}
\begin{figure}
\centering
\includegraphics[width=.50\textwidth]{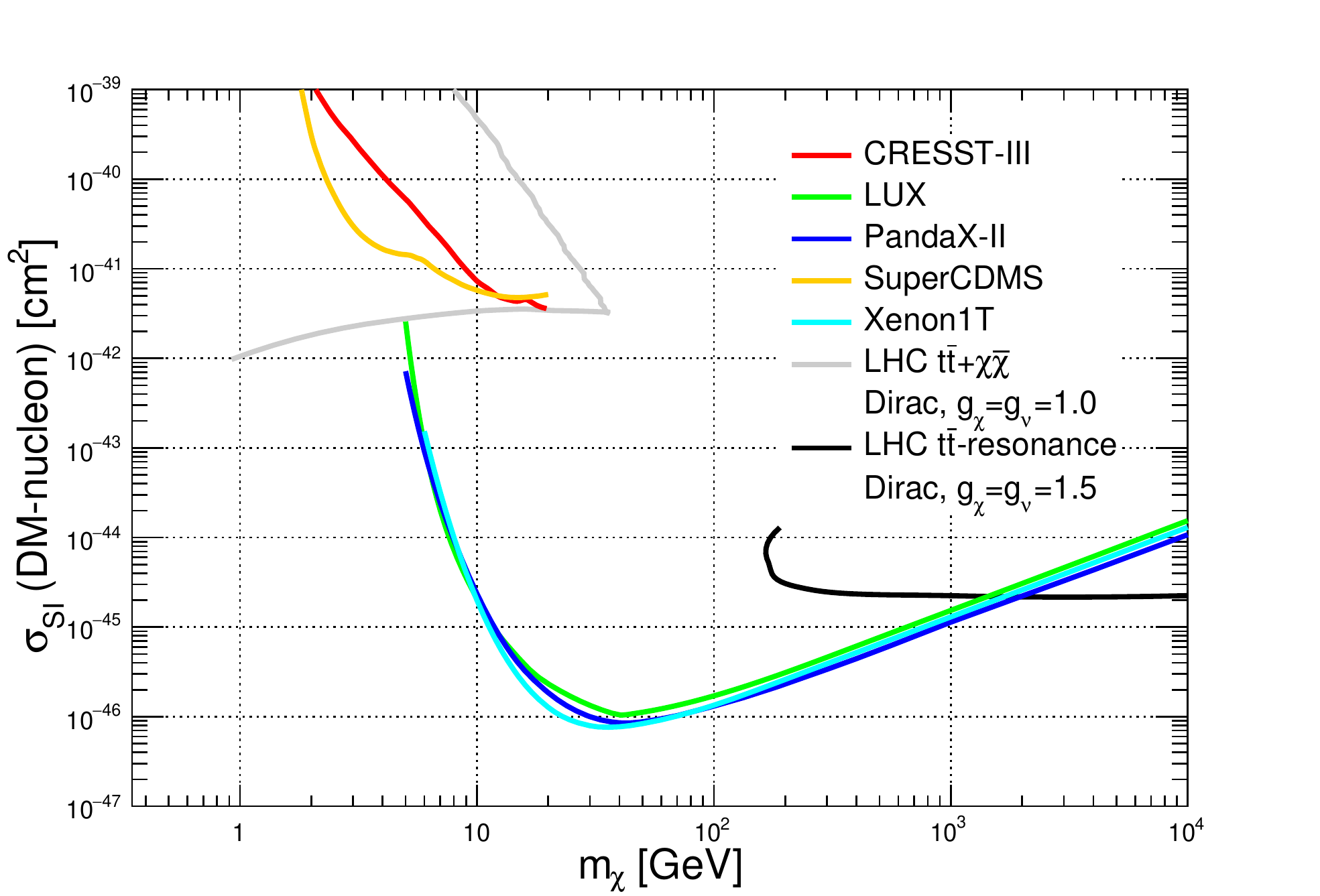}

\caption{Upper limit on spin-independent DM-nucleon cross section ($\sigma_{SI}$) as a function of the DM mass.
The exclusion limit of the $t \bar{t}$ resonance obtained in this paper at 95\% CL, is compared with limits from direct detection experiments and from the latest $t \bar{t} + \chi \bar{\chi}$ by ATLAS at 90\% CL.
}
\centering
\label{fig:DD}
\end{figure}

For scalar simplified models, we set an upper limit on the DM-nucleon cross section ($\sigma_{SI}$) as follows \cite{Boveia:2016mrp}:
\begin{equation}
\sigma_{SI} \simeq 6.9 \cdot 10^{-43} \cdot cm^2 \cdot (g_\nu \cdot g_\chi)^2 \Big(\frac{125 GeV}{m_\phi}\Big)^4 \Big(\frac{\mu_{n \chi}}{1 \mbox{GeV}}\Big)^2
\label{eq:DD_limits},
\end{equation} 
where we introduce the nucleon mass, $m_n = 0.939$ GeV (for both Neutrons and Protons), and the reduced DM-nucleon mass, $m_n \cdot m_\chi / (m_n + m_\chi)$.
In the results presented at \ref{sec:results_simplified}, we find that scalar masses in the range of $m_\phi \in [400,600]$ GeV are excluded, with the lowest couplings in the range of $g \in [1.1-2.0]$ and $m_\chi \geq 160 GeV$.
In order to use the prescription above, the couplings should be fixed. Therefore we choose $g_\chi = g_v = 1.5$.
The results obtained in this paper are presented at Figure \ref{fig:DD}, along with direct detection experiments \cite{Akerib:2016vxi,Ren:2018gyx,Aprile:2017iyp,Agnese:2015nto,Petricca:2017zdp} and with the contour calculated at \cite{Aaboud:2017rzf}.
There are two caveats for this comparison, however: the results we state are at 95\% CL, while the results of the other experiments are at 90\% CL; furthermore, one should keep in mind that the comparison is model dependent.

\subsubsection{Spin-Dependent}
For the pseudo-scalar scenario, a velocity suppression term in the non-relativistic limit creates large difference by several orders of magnitude in favour of LHC results \cite{Aaboud:2017rzf, Boveia:2016mrp}, and therefore it is not presented.
However, this actually makes the motivation of using $t \bar{t}$ resonance interpretation for DM models even stronger: it covers regions with higher DM masses, for which both $t \bar{t} + \chi \bar{\chi}$ and direct detection searches are insensitive to.

\section{Conclusion}
\label{sec:conclusion}
The necessity of $t \bar{t}$ resonance search as a complementary measurement for models containing DM candidates has been emphasized for two different models.
Despite the challenges of this signature  from the experimental point of view, it provides unique access to regions in the parameter space with high DM candidates masses, where other signatures quite often do not have the sensitivity. 
Further expansion of the parameter space covered by this signature is expected, using centre of mass energy of $\sqrt{s}=13$ TeV, and with the higher luminosity now recorded by ATLAS and CMS collaborations.

\begin{acknowledgements}
We thank Jonathan Cohen for the help with the theoretical part of the paper, for reviewing the paper, and for many useful and enlightening discussions.
We thank Priscilla Pani for reviewing the paper and sharing here enlightening comments with us.
This research was supported by a grant from the United States-Israel Binational Science Foundation (BSF), Jerusalem, Israel, and by a grant from the Israel Science Foundation (ISF).
\end{acknowledgements}

\end{document}